# Spin-group symmetry in magnetic materials with negligible spin-orbit coupling

Pengfei Liu[1,2], Jiayu Li[1], Jingzhi Han[2], Xiangang Wan[3,*] and Qihang Liu[1,4,5,*]

[1]*Shenzhen Institute for Quantum Science and Engineering and Department of Physics, Southern University of Science and Technology, Shenzhen 518055, China*

[2]*School of Physics, Peking University, Beijing 100871, People's Republic of China*

[3]*National Laboratory of Solid State Microstructures and School of Physics, Nanjing University, Nanjing 210093, China and Collaborative Innovation Center of Advanced Microstructures, Nanjing University, Nanjing 210093, China*

[4]*Guangdong Provincial Key Laboratory of Computational Science and Material Design, Southern University of Science and Technology, Shenzhen 518055, China*

[5]*Shenzhen Key Laboratory of Advanced Quantum Functional Materials and Devices, Southern University of Science and Technology, Shenzhen 518055, China*

[*]Emails: xgwan@nju.edu.cn; liuqh@sustech.edu.cn

**Abstract**

Symmetry formulated by group theory plays an essential role with respect to the laws of nature, from fundamental particles to condensed matter systems. Here, by combining symmetry analysis and model calculations, we elucidate that the crystallographic symmetries of a vast number of magnetic materials with light elements, in which the neglect of relativistic spin-orbit coupling (SOC) is an appropriate approximation, are considerably larger than the conventional magnetic groups. Thus, a symmetry description that involves partially-decoupled spin and spatial rotations, dubbed as spin group, is required. We derive the classifications of spin point groups describing coplanar and collinear magnetic structures, and the irreducible co-representations of spin space groups illustrating more energy degeneracies that are disallowed by magnetic groups. One consequence of the spin group is the new anti-unitary symmetries that protect SOC-free $Z_2$ topological phases with unprecedented surface node structures. Our work not only manifests the physical reality of materials with weak SOC, but also sheds light on the understanding of all solids with and without SOC by a unified group theory.

**I. Introduction**

The study of symmetry has always been the kernel of condensed matter physics and materials chemistry, as it dictates the way in which wavefunctions of elementary excitations behave, including geometric phases, selection rules, and degeneracies. The corresponding wavefunction properties thus reflect on the physical observables such as polarization, response susceptibility, and band dispersions. The symmetries of three-dimensional (3D) solids are believed to be described by a complete crystallographic group theory, including 32 point groups (PGs), 230 space groups (SGs), 122 magnetic PGs (MPGs), 1651 magnetic SGs (MSGs), and their double groups with spinor representations [1,2]. They apply to the nonmagnetic materials without and with spin-orbit coupling (SG and double SG) as well as the magnetic materials with spin-orbit coupling (double MSG). The recent prosperity of symmetry-protected topological phase in condensed matter systems is based on the electronic structures, providing a fertile playground for a survey of various quasi-particles including Weyl, Dirac fermions, and others beyond them [3-15]. Moreover, the theories of symmetry-based indicator and (magnetic) topological chemistry based on band representations allow a comprehensive classification of topological crystalline insulators and semimetals, leading to a dictionary of thousands of predicted topological materials [16-31]. Notably, topological phases without spin-orbit coupling (SOC) are also widely investigated in nonmagnetic materials as a starting point of the Hamiltonian with relativistic SOC [13,32-34].

However, the symmetry description of the remaining quadrant, i.e., the magnetic materials with negligible SOC, which represents a vast number of compounds with light elements, is much less explored (see Fig. 1). The most striking characteristic of the materials in this quadrant is the nontrivial spin degrees of freedom yet decoupled with the orbital part. Thus, the corresponding symmetry is not fully described by any of the abovementioned SGs. Specifically, the symmetry operations of spin, e.g., spin rotations and the symmetry operations of lattice, can combine in the way disallowed by SOC, which form a composite symmetry group applied on both position space and spin space but not necessarily simultaneously. Such groups, dubbed as spin group including spin point group (SPG) and spin space group (SSG), was first considered in 1960-1970s to account for the extra symmetries of Heisenberg Hamiltonian with the application on spin waves [35,36]. Later, spin group is defined in a mathematically rigorous way, where all of the 598 nontrivial spin point groups (in which pure spin operations are excluded) are classified using an analogous method of classifying 90 type-I and type-III crystallographic MPGs [37,38]. Apart from the original works, there are scattered works discussing the applications of spin group symmetry in Landau theory of phase transition [39,40], neutron scattering [41], electronic states with spiral magnetic order [42], etc.

As a complete characterization of symmetry in magnetic materials when SOC is negligible, the theory of spin group is expected to be powerful also in the application of electronic ground-state properties such as spin/orbit polarization, Berry curvature, and linear responses such as anomalous/spin Hall conductance, (inverse) spin Galvanic effect, (inverse) Faraday effect, etc. Surprisingly, such topics are seldomly explored until the recent growing interest of antiferromagnetic (AFM) spintronics [43,44]. It is found that certain effects in AFM materials could appear without the assistance of SOC, including transverse spin current in collinear [45-47] and non-collinear AFM systems [48-50], AFM-induced spin splitting [51-56], giant piezomagnetism [47], etc. In [48,49], spin group symmetry is used for determining the conductivity tensor of $Mn_3X$ (X = Sn, Ir, Ga, and Ge) without SOC, showing more symmetry restrictions for allowed spin conductivity compared to the case with SOC. In [55,56], broken of a combined operation of translation and pure spin rotation that flip spin direction gives rise to a type of antiferromagnets with spin-split bands. In the other works concerning nontrivial effects of AFM order, spin group symmetry may not be taken into account but actually can act as an effective tool in explaining the predicted results.

Indeed, it is aware that symmetries concerning spin and orbital degrees of freedom are decoupled in magnetic materials with negligible SOC, but such symmetries can hardly be fully comprehended without spin group theory. Most previous studies related to spin group symmetries consider symmetry operations leaving a specific Hamiltonian invariant, posing limitations in applying a uniform theory to magnetic materials with all kinds of spin arrangements, including collinear, coplanar, and noncoplanar spin configurations. In [57], for example, the Hamiltonians considering collinear magnetic order are block-diagonal, which greatly simplifies the symmetry analysis of the system. However, such an approach is not applicable to non-collinear magnetic configurations even without SOC. On the other hand, the recent progress of condensed matter physics, in which the geometric phase, topological matter, and emergent quasi-particles play an essential role, paves an avenue for the application of spin groups in describing complicated magnetic phases. Hence, there is an urgent need to establish the connection between the powerful spin group theory and the frontier of modern condensed matter studies.

Here, we systematically study the description of spin-½ electrons in magnetic lattices by spin groups and their applications on the topological electronic structures. Starting from general single-electron Hamiltonians, we first demonstrate that the description of the full symmetry group of a spin-orbit-decoupled system with on-site local magnetic moments naturally points to spin groups (Sec. II). Exemplified by a spinful hexagonal molecule model, we pedagogically elucidate that compared with magnetic groups where the spin and spatial rotations are completely locked to each other, in spin groups, more discrete or continuous spin rotations under a spatial rotation are permitted. We further derive comprehensive classifications of SPGs that are direct products of the nontrivial and spin-only part, including 252 and 90 SPGs for describing the coplanar and collinear magnetic structures, respectively (Sec. III). Exemplified by a kagome lattice model, we derive the irreducible co-representations of the little spin group at high-symmetry momenta and illustrate the resulting energy degeneracies that cannot be understood by the conventional magnetic (double) groups (Sec. IV). Within the regime of spin group, we find more symmetry operations that would lead to new topological phases such as SOC-free $Z_2$ topological insulator (TI) with unprecedented surface node structures, further enriching the existing zoo of the topological materials (Sec. V). For materials realization, we show that square-net compounds $AMnBi_2$ (A = Sr, Ca) could realize such $Z_2$ topological phases with surface nodal lines as well as bulk Dirac points at generic momenta protected by spin group symmetries (Sec. VI). Our work appeals to the general interest in utilizing spin group in various fields with an expanded material pool for potential spintronics applications, which contains candidates with both light elements and nontrivial electronic properties.

**II. General single-electron Hamiltonians**

To begin with, we apply spatial and spin rotational operations on the general steady-state Hamiltonians to illustrate the requirements of symmetry operations in various cases. For a nonmagnetic system without SOC [ $H = \frac{\hat{p}^2}{2m} + V(\hat{r})$ ], the wavefunctions are labeled by the single-valued representations of the $V(\hat{r})$-determined SG, of which the rotational elements contain the spatial rotations $C_n(\theta)$ solely. When the general SOC term $H_{soc} = \frac{1}{2m^2c^2}(\nabla V(\hat{r}) \times \hat{p}) \cdot \hat{\sigma}$ is added, neither spatial rotation $C_n(\theta)$ nor spin rotation $U_m(\varphi)$ alone, but only a locked combination $U_n(\theta)C_n(\theta)$ can keep $H_{soc}$ invariant. The resulting spinor wavefunctions furnish the double-valued representations of the $V(\hat{r})$-determined double SG. To describe the magnetic systems, we apply $H_{mag} = S(\hat{r}) \cdot \hat{\sigma}$ under the framework of single-particle mean-field approximation [58-60], where $S(\hat{r})$ stands for the $r$-dependent exchange field due to the distribution of local magnetic moments and $\hat{\sigma}$ is, strictly speaking, not the spin of electrons but the spin of quasi-particles arising from exchange-correlation among electrons. Note that $H_{mag}$ resembles the form of double-exchange model widely used to describe the magnetic phase transitions of manganites [61-63]. If $H + H_{soc} + H_{mag}$ is considered, the locking between spin and spatial rotations still holds, while the full symmetry description requires double MSG with the inclusion of time-reversal operation $T$.

If we consider the symmetry operations of $H + H_{mag}$, i.e., a magnetic system without SOC, it is straightforward to prove that the complete locking of spatial and spin rotations is no longer required. Instead, partially locked rotations, i.e., $U_m(\varphi)C_n(\theta)$, with $C_n(\theta)$ keeping $V(\hat{r})$ invariant, could keep $H_{mag}$ invariant and thus the total Hamiltonian. Note that $\boldsymbol{S}(\hat{\boldsymbol{r}})$ presents the effects of spatially distributed magnetic moments residing at the atomic sites with a given magnetization direction, coupling with the spin through exchange-correlation interactions. Unlike SOC, such a spin-spatial coupling is nonrelativistic and constrain $U_m(\varphi)$ with $C_n(\theta)$ in various ways according to the specific spin arrangement, forming spin groups, as will be discussed below (see Appendix A-C). The Hamiltonians and symmetry groups describing different systems are summarized in Fig. 1, with the derivation of the constraints provided in Supplementary Section II [64].

**III. Spin point group: partially decoupled spin and spatial rotation**

Spin group includes SPG and SSG. We first discuss SPG, whose elements are denoted by $\{U_m(\varphi), TU_m(\varphi)||C_n(\theta), IC_n(\theta)\}$. While spatial inversion $I$ can combine $C_n(\theta)$ forming improper rotations and mirror reflections, time-reversal symmetry $T$ reverses spin and is thus written in the spin space ($T^2 = -1$ in spin-1/2 electron system) [72]. For simplicity, we first consider proper rotations only, i.e., $\{U_m(\varphi)||C_n(\theta)\}$. The partial coupling between spin and spatial operations implies pure spin rotation $\{U_m(\varphi)||E\}$ and coupled spin-spatial rotation, the latter of which forms nontrivial SPGs containing elements of the form $\{U_m(\varphi)||C_n(\theta)\}$ with $C_n(\theta) \neq E$ ($E$ is identity rotation) except identity element $\{E||E\}$. Ref. [38] constructed all nontrivial SPGs in 3D crystals, by combining the factor groups of PGs and their exhaustive isomorphic groups as the spin part. Here, differ from mathematical construction, we focus on an exemplified structure with spin arrangements to illustrate how does the regime of spin group differentiate conventional magnetic group in permitting much more symmetry operations and the resulting physical consequence in a magnetic material with negligible SOC in terms of band degeneracy and topological electronic structure.

To illustrate that for certain configurations, SPG generally possesses more symmetry operations than the conventional MPG, and we consider a spinful hexagonal molecular structure with the spatial rotational group $D_6$ with generators $C_{6z}$ ($C_z(\frac{\pi}{3})$) and $C_{2x}$ ($C_x(\pi)$), as shown in Fig. 2(a)-2(h). Placing magnetic moments on each site, in general, reduces the $D_6$ symmetry. Considering MPG symmetry (with SOC), the only spin configuration that maintains $D_6$ symmetry is the in-plane spin arrangement shown in Fig. 2(c), while there are more possibilities for SPG symmetry. Without the loss of generality, we build a single-orbital (e.g., $d_{z^2}$) tight-binding (TB) model with in-plane local magnetic moments having the same magnitude but different coplanar directions, $\boldsymbol{S}_i = S[Cos(\phi_i), Sin(\phi_i), 0]$ (see Fig. 2(a)). The matrix elements of the Hamiltonian are written as

$$\langle i, \boldsymbol{z}_i \cdot \boldsymbol{\sigma} = \alpha | \hat{H} | j, \boldsymbol{z}_j \cdot \boldsymbol{\sigma} = \beta \rangle = t[\delta_{i,j+1} + \delta_{i,j-1}] X(\phi_i - \phi_j)_{\alpha\beta} + \delta_{i,j}(JS\sigma_z)_{\alpha\beta}, \quad (1)$$

where we choose the basis functions with local spin quantization axis directing along local magnetic moments, i.e., $\boldsymbol{z}_i = [Cos(\phi_i), Sin(\phi_i), 0]$, and $X(\theta)$ is defined as $X(\theta) \equiv \begin{bmatrix} Cos\left(\frac{\theta}{2}\right) & i\,Sin\left(\frac{\theta}{2}\right) \\ i\,Sin\left(\frac{\theta}{2}\right) & Cos\left(\frac{\theta}{2}\right) \end{bmatrix}$. We then check all the possible $\{U_m(\varphi)||C_z(\frac{\pi}{3})\}$ and $\{U_m(\varphi)||C_x(\pi)\}$ operations that leave the Hamiltonian invariant and find that a spatial rotation $C_n\left(\frac{2\pi}{d}\right)$ could couple a spin rotation $U_m\left(\frac{2p\pi}{d}\right)$ with $d$ being the order of rotation and $p = 0, 1, \ldots d - 1$ (see Supplementary Section III [64]). Consequently, there are 7 inequivalent types of spin configurations containing $C_z(\frac{\pi}{3})$ and $C_x(\pi)$ spatial rotations, including one collinear ferromagnetic (FM), one collinear AFM, one non-collinear FM, and 4 coplanar AFM configurations. Their nontrivial SPG symbols and generators are shown in Fig. 2(b)-2(h). Furthermore, the matrix elements of the spin-group Hamiltonian are functions of the angles between the local moments of the neighboring sites; hence, rotating all the moments by the same angle leaves the eigenvalues of the Hamiltonian invariant indicating the decoupling between spin space and real space. The abovementioned properties explicitly elucidate how do the spin and spatial rotation "partially" couple to each other in magnetic materials without SOC.

By considering spatial and spin rotation separately, the SOC effect could be considered as a constraint to limit the relationship of $C_n(\theta)$ and $U_m(\varphi)$ that reduce symmetry. Consequently, spin group itself could serve as a unified theory of both nonmagnetic and magnetic groups, with and without SOC. We summarize the symmetry hierarchy in the context of SPG operations in Fig. 2(i). The specific hierarchy diagram for the spinful hexagonal molecule with various spin arrangements is shown in Supplementary Section III [64]. The nonmagnetic or paramagnetic phase without SOC, where spin rotation is fully independent of the spatial operations, has the highest symmetry, i.e., the direct product of the spatial part $G_{p0}$ and spin part $SO(3) \times Z_2^T$ ($Z_2^T = \{E, T\}$, "×" denotes the internal direct product, while "⊗" denotes (external) direct product of two groups, see Supplementary Section I [64]). With SOC, the symmetry degrades to subgroups $G_p \times Z_2^T$ by adding the constraint of the complete spin-space coupling (type II MPG, 32 grey groups). The further addition of magnetic orders leads to conventional magnetic group $G_{mp}$, including type I (32 colorless MPGs) and type III (58 black-white MPGs).

The symmetry hierarchy also has another branch by adding magnetic order first and then SOC, leading to SPGs and MPGs, respectively. There are 598 nontrivial SPGs $G_{nsp}$[38], which could describe noncoplanar spin arrangements. In addition, for coplanar moments, there exists a boson-like time-reversal group that forms the spin-only PG $Z_2^K = \{E, TU_m(\pi) = K\}$, where $K$ denotes complex conjugation, rendering the full SPG $G_{sp} = G_{nsp} \times Z_2^K$. For collinear moments, the full SPGs is $G_{sp} = G_{nsp} \times (SO(2) \rtimes Z_2^K)$ with an additional $SO(2)$ rotational symmetry group along the common direction of the spins (see Appendix B). After a comprehensive classification, we obtain 252 and 90 SPGs for describing the symmetries of coplanar and collinear magnetic structures without SOC, respectively (see Appendix C). The physical effects of $TU_m(\pi)$ symmetry in coplanar and collinear magnets will be discussed later.

### IV. Spin space group and band degeneracy

By considering the translational symmetry of the lattice, one can easily generalize the operations of SPG to SSG, $\{U_m(\varphi)||C_n(\theta)|\tau\}$, where $\tau$ denotes the spatial translation within a primitive cell. Similarly, the analogous symmetry hierarchy for SPG in Fig. 2(i) could be generalized to SSG by involving color Bravais lattices, which certainly includes the current 1651 MSGs (also known as Shubnikov groups). Because of the complexity of color Bravais lattices, the exhaustive construction of SSGs is complicated, with an infinite number of possible types.

We next perform a case study to illustrate the additional band degeneracies induced by SSG symmetry. We consider a transition-metal layer with kagome lattice and non-collinear AFM spin configuration shown in Fig. 2(e), which is similar to the spin arrangement of the non-collinear antiferromagnets Mn$_3$Ge and Mn$_3$Sn [73-76], as shown in Fig. 3(a). By constructing a simple single-orbital TB lattice model, we show that the electronic structure of such a magnetic lattice system should be described by SSG rather than MSG. The lattice Hamiltonian is written as follows:

$$H = \sum_{\alpha,\beta}\left(\sum_{<R,i;R',j>} t\, a^\dagger_{R,i,\alpha} \delta_{\alpha,\beta} a_{R',j,\beta} + J \sum_{R,\,i} a^\dagger_{R,i,\alpha} (\boldsymbol{S_i} \cdot \boldsymbol{\sigma})_{\alpha,\beta} a_{R,\,i,\,\beta}\right). \quad (2)$$

The first term is the nearest neighbor hopping, and the second term counts the effect of the local magnetic moment $\boldsymbol{S_i}$. Fig. 3(b) and 3(c) show the SOC-free band structures of such a 6-band model (including spin) without and with magnetic order, respectively. The nonmagnetic kagome structure with P6/mmm symmetry exhibits its prototypical band structure, including a flat band and a Dirac cone at K. By adding the noncolinear AFM order, the spin-degenerate bands in Fig. 3(b) splits to two sets of Dirac cones without opening a gap. Apparently, such two-fold degeneracy certainly cannot be interpreted with a MSG $Cmm'm'$ that only supports 1D irreducible co-representations at the K point. Such degeneracies were also obtained in band structures of non-collinear AFM kagome models [49,77]. While for collinear AFM order, the Hamiltonian is block-diagonal for spin-up and spin-down components (in this case, each block can be viewed as a spinless system with $T^2 = 1$) [57,78], the noncolinear Hamiltonian such as Eq. (2) cannot be decomposed straightforwardly. Hence, we apply the group representation theory that fully describes the system to analyze the band degeneracies at the high-symmetry momenta.

First of all, we construct all irreducible co-representations of the little spin group at a high-symmetry momentum via the similar procedure of deriving irreducible co-representations for magnetic groups [2]. Thus, the compatibility relations, i.e., correspondence among representations of a nonmagnetic group, spin group, and magnetic group, could be obtained (see Appendix D for the methods of obtaining irreducible co-representations of little co-groups). Then, starting from the representations of the nonmagnetic site groups furnished by atomic orbitals sitting on a certain Wyckoff position, we obtain the nonmagnetic band representations of the little group at a high-symmetry momentum [16]. Finally, the correspondence between single-valued nonmagnetic representations and spin group co-representations gives rise to the resulting band representations of a spin group (see Supplementary Section V for the details of the TB model as well as the irreducible co-representations of little co-groups [64]).

The band degeneracy and compatibility relations at the $K$ valley, shown in Fig. 3(c), can be successfully explained by spin group symmetry. As shown in Fig. 3(b), the $d_{z^2}$ orbitals on the three kagome sites (Wyckoff position $3g$) give rise to two kinds of representations $K_1(2)$ and $K_5(4)$ including spin (the number in parentheses denotes the degree of degeneracy). By choosing the eigenstates of spatial rotoinversion $C_{6z}I$ as the basis functions and adding spin degrees of freedom, we operate the spin-group generators $\{U_z(\frac{2\pi}{3})||C_z(\frac{\pi}{3})I|0\}$, $\{U_x(\pi)||C_x(\pi)|0\}$ and $\{TU_z(\pi)||I|0\}$ on the Hamiltonian and find that $K_5(4)$ splits to three levels with two 1D irreducible co-representation $K_1^s$ (1), $K_2^s$ (1), and a 2D irreducible co-representation $K_3^s$ (2), consistent with Fig. 3(c) calculated by the TB model. The compatibility of the band splitting is summarized in Fig. 3 (d).

**V. Quasi-Kramers degeneracy and $Z_2$ magnetic topological insulator**

Spin-group operations, including an enormous number of combinations of pure spatial operations, time-reversal, and pure spin rotations, significantly enhance the symmetries of magnetic materials. As a result, there have to be various extra topological phases unexpected before, protected by spin-group symmetries. Analogous to $Z_2$ TIs protected by $T$, we next consider anti-unitary operations squared into −1 in spin groups, which could allow $Z_2$ topological classification in 2D subspaces of the 3D Brillouin zone [79]. Note that there are other $Z_2$ topological classifications protected by crystalline symmetries that are unitary or squared to 1 [80-82], which can also be extended to spin group symmetries. We list all such symmetry operations in the regime of SSG in Table I and define the resulting degeneracy at certain high-symmetry *k*-points as "quasi-Kramers degeneracy". Furthermore, we only consider the symmetries that still persist in certain cleaved surfaces, i.e., possibly having symmetry-protected topological surface states (e.g., $\{T||I|0\}$ and $\{TU_n(\pi)||C_z(\pi)|\tau_{z/2}\}$ are excluded).

It is found that only two symmetry operations in magnetic materials also exist in MSG, corresponding to AFM $Z_2$ TI (e.g., MnBi$_2$Te$_4$) [83-87] and topological semimetals with symmetry-protected double helicoid surface states predicted in nonmagnetic systems [88,89], respectively. We confirm that such topological phases could still exist without SOC. On the other hand, the other five symmetries supporting quasi-Kramers degeneracy exist solely in SSGs, without any analogues in nonmagnetic materials or magnetic materials with large SOC. Among them, $\{T||C_z(\pi)|0\}$, $\{T||m_{[001]}|0\}$, $\{T||m_{[001]}|\tau_{x/2}\}$ contain both pure spatial rotations, and the spin-1/2 time-reversal $T$, and are thus spin-group symmetries because of the decoupling between spin and lattice. The (001) surface bands with $\{T||C_z(\pi)|0\}$ symmetry are all doubly degenerate, and thus do not protect gapless surface states in general. However, if the **z**-axis is the axis of high rotational symmetry or there are additional spin rotational symmetries, the surface states may manifest double Dirac point, which is predicted only in bulk bands before [10,90]. The three spin-group symmetries containing spatial mirror reflection $m_z$ supports surface Dirac nodal line, which is not reported in magnetic systems before. Among them, $\{T||m_{[001]}|0\}$ could lead to $Z_2$ classification for a system and the corresponding 3D quantum spin Hall phases. The last symmetry $\{TU_n(\pi)||E|\tau_{z/2}\}$ also supports $Z_2$ magnetic TI, similar to $\{T||E|\tau_{z/2}\}$, except that the surface Dirac point is located at $(0, \pi)$ or $(\pi, \pi)$ momenta in the momentum space. We note that similar emergent $Z_2$ topological phases protected by order-2 nonsymmorphic antiunitary symmetries were discussed mathematically [91], while $\{TU_n(\pi)||E|\tau_{z/2}\}$ in SSG provides a physical scenario to realize such exotic topological phases.

We next take $\{T||m_{[001]}|0\}$ and $\{TU_n(\pi)||E|\tau_{z/2}\}$ as examples to illustrate the new $Z_2$ magnetic TIs and various unexpected surface node structures. We start from a 3D Dirac semimetal model without SOC [70], which is analogous to a 3D version of graphene. Such a phase can easily transform to a Weyl semimetal under local magnetic moments $S_z$ along the $z$ direction. We can thus tune the hopping parameters to realize a Chern insulator phase at $k_z = \pi/2$ plane of the Brillion zone. Then, by building an AFM structure through cell-doubling (Fig. 4(a)), we can annihilate the Weyl points with opposite chirality and create a gapped insulator. By constructing an 8-band model (see Supplementary Section VI for details [64]), we realize a $Z_2$ magnetic TI protected by both $\{T||m_{[001]}|0\}$ and $\{TU_n(\pi)||E|\tau_{z/2}\}$ symmetries, with gapless Dirac surface states at the boundaries of all 2D planes perpendicular to the $k_z$ direction. Consequently, it manifests surface Dirac node lines at $k_x = 0$ or $k_x = \pi$ line for any surfaces perpendicular to $m_{[001]}$ (Fig. 4(c) and 4(e)), which is impossible in conventional TIs protected by $T$ or $\{T||E|\tau_{z/2}\}$. To examine the impact of each symmetry, we apply an in-plane FM canting to break $\{T||m_{[001]}|0\}$ (Fig. 4(b)), then the surface node structure becomes a Dirac point at $(0, \pi)$ (Fig. 4(d) and 4(f)), which is consistent with the symmetry analysis shown in Table I. If $\{TU_n(\pi)||E|\tau_{z/2}\}$ is broken by the dimerization of the two layers, the Dirac point at $(0, \pi)$ is finally gapped (Fig. 4(g) and 4(h)). Therefore, we demonstrate that unlike nonmagnetic materials and magnetic materials with SOC, magnetic materials with negligible SOC possess new $Z_2$ topological classification with unprecedented surface node structures protected by SSG symmetries. We note that the previous studies about SOC-free TIs focused on spinless systems protected by pure crystalline symmetry without considering spin rotation or time-reversal symmetry, which differs from the situation discussed here [32].

**VI. Materials realization of nodal-line semimetals**

A remarkable consequence of symmetry and topology in the electronic structure of materials is the existence of protected degeneracies, leading to various topological semimetals such as Dirac, Weyl, nodal-line, and nodal-surface semimetals. Thus, the corresponding symmetry design principles could be established to conduct a comprehensive material search. Here we take two widely studied magnetic topological semimetals, e.g., SrMnBi$_2$ and CaMnBi$_2$, to illustrate their unrevealed bulk and surface nodes that only exist under the regime of spin-group symmetries, including Dirac points at arbitrary *k*-points and surface nodal lines. Note that we turn off SOC in the calculation of these well-studied large-SOC materials to illustrate the distinct topological phases in a semi-realistic setup. The identification of more suitable material candidates described by spin-group symmetries is left for future works.

AMnBi$_2$ (A = Sr, Ca) are layered materials with anisotropic Dirac fermions (see Fig. 5(a)), inspiring the study of square-net materials as topological semimetals such as ZrSiS [92-94]. Despite a checkerboard-type AFM configuration, the nodal properties in such compounds are typically treated via nonmagnetic models [71]. Without SOC, the diagnosis for nonmagnetic topological semimetals with inversion symmetry indicates that AMnBi$_2$ are nodal-line semimetals [34] (see Supplementary Section VII [64]). The AFM magnetic order further turns the nodal lines to discrete Dirac points with four-fold degeneracy. Compared with the case with SOC where the Dirac points only occur at high-order rotational axes or Brillouin zone boundary, the Dirac points protected by spin-group symmetry could occur even at arbitrary *k*-points, like chiral Weyl points. Such peculiar property could be understood by the $k \cdot p$ low-energy Hamiltonian. By applying $\{T||I|0\}$ ($i\sigma_y \tau_x K$) and $SO(2)$ spin rotation along the spin direction ($e^{-i\theta\sigma_z}$), the symmetry-allowed Hamiltonian takes the form $H(\boldsymbol{k}) =$ $f_0(\mathbf{k})\sigma_0\tau_0 + f_1(\mathbf{k})\sigma_0\tau_x + f_2(\mathbf{k})\sigma_0\tau_y + f_3(\mathbf{k})\sigma_z\tau_z$ with the basis $\{|\psi_A \uparrow\rangle, |\psi_B \uparrow\rangle, |\psi_A \downarrow\rangle, |\psi_B \downarrow\rangle\}$, where A and B represent two sublattices connected by $\{T||I|0\}$. The last three terms of $H(\mathbf{k})$ mutually anti-commute with each other, leading to stable Dirac nodes that could appear at generic momenta and cannot be gapped by any perturbation that maintain $\{T||I|0\}$, $SO(2)$ spin rotation and translation symmetries. The Dirac points of SrMnBi$_2$ calculated by density functional theory (DFT) are shown in Fig. 5(b) (see methods of DFT calculations in Appendix E).

According to the above discussion, $\{T||m_{[110]}|0\}$ spin-group symmetry in AMnBi$_2$ protects $Z_2$ topological classification with unprecedented surface nodal lines. We next apply DFT calculations on SrMnBi$_2$ under uniaxial pressure (the lattice parameter $c$ (along the z-direction) is reduced by 10%) to verify this. Fig. 5(c) plots the surface states on the (001) surface, showing gapless Dirac cone at both $\bar{\Gamma}'$ and $\bar{M}'$ point. The existence of the two Dirac points at $(0.280, 0.280, 0)$ and $(0.293, 0.293, 0.272)$, protects a region in which any vertical planes in the Brillouin zone parallel to $m_{[110]}$ yield a nontrivial 2D $Z_2 = 1$ phase, as indicated by the Wilson loop of a representative plane (shown in green in Fig. 5(b)) and the transition of $Z_2$ as a function of the momentum along the [110] direction (see Figs. 5(d) and 5(e)). Furthermore, the surface nodes form a line between the surface projections of the two Dirac points. The four-fold rotation symmetry in this system transforms the nodal line into 4 nodal lines, with two protected by $\{T||m_{[110]}|0\}$, and the other two protected by $\{T||m_{[1\bar{1}0]}|0\}$.

## VII. Discussion

Although established decades ago, the concept of spin group is not widely explored or applied due to the lack of suitable condensed matter scenarios, especially in spin-½ electronic systems. The main purpose of our work is to build the bridge between the powerful but largely overlooked symmetry group and the frontier of quantum material studies. The abovementioned symmetry-protected degeneracy and $Z_2$ topological classification are merely the tip of the iceberg for the application of the spin group, leaving fruitful diversity of topological phases and emergent fermions induced by such an enhanced symmetry group to be further explored. For example, a recent work realizes a condensed-matter counterpart of the *SU*(2) flavor symmetry in particle physics, leading to a new type of AFM Weyl semimetal protected by spin group symmetries [95]. Such nodal structures in both bulk and surface states could also shed light on the non-Abelian band topology in magnetic metals [13]. Furthermore, symmetry indicators based on spin group would also give rise to more possibilities of topological crystalline insulators and semimetals.

We next discuss two physical effects of the spin mirror symmetry $\{TU_z(\pi)||E|0\}$ existing in coplanar and collinear spin groups (the comprehensive list of 252 and 90 SPGs, respectively, is provided in Supplementary Sec. IV [64]). The first one is spin-momentum locking. For coplanar spin groups, if spin-splitting exists owing to broken $\{T||I|0\}$ symmetry, the spin texture, i.e., the expectation of spin operator with respect to the momentum $S(\mathbf{k})$, has to be $(S_x(\mathbf{k}), S_y(\mathbf{k}), S_z(\mathbf{k})) = (S_x(-\mathbf{k}), S_y(-\mathbf{k}), -S_z(-\mathbf{k}))$. In addition, for collinear spin groups with the moments along the *x*-axis, the existence of SO(2) group forces nondegenerate energy eigenstates to be the eigenstates of $\sigma_x$. Hence, only inversion symmetric $S_x$ components survive, i.e., $(S_x(\mathbf{k}), 0, 0) = (S_x(-\mathbf{k}), 0, 0)$. The second one is charge and spin transports. Because the expression of Berry curvature is independent of spin, for coplanar magnets, $\{TU_z(\pi)||E|0\}$ forces the integral of Berry curvature throughout the Brilloun zone to be zero, in analogy to the role of *T* for nonmagnetic systems. As a result, anomalous Hall effect cannot be induced by coplanar magnetic orders without the assistance of perturbations such as SOC and magnetic tilting. However, $\{TU_z(\pi)||E|0\}$, in general, does not forbid transverse spin current and spin Hall effect in both coplanar [48-50] and collinear AFM systems [45-47]. Particularly, in coplanar AFM systems, both *T*-odd (transverse spin current) and *T*-even (spin Hall effect) contributions could be nonzero [48,49], while only *T*-odd contribution exists in collinear AFM systems.

While SOC is an intrinsic relativistic property for all materials depending on the atomic mass of the constituting elements, the theory of spin-group, which describes the symmetry of a magnetic ground state, acts as the very starting point to understand the behavior of magnetic materials with SOC. For most materials, even with strong

SOC, e.g., 10-100 meV, its influence on the electronic structure is still small compared with those caused by hopping, exchange splitting, crystal field, etc. (typically in the order of eV). Consequently, one can construct the zero-order Hamiltonian of a magnetic ground state based on spin group and add SOC as high-order perturbation terms. Such an approach also provides an alternative paradigm to accurately understand the role of SOC by differentiating the contribution of SOC and the contribution of magnetic moments and crystal lattice. In addition, each element of the response tensors $\chi^{(n)}$ in Kubo formalism for observables like spin-orbit torque correlates the symmetry of crystals [96], determining its zero/nonzero value but not the magnitude. Therefore, the conventional MSG cannot tell if a symmetry-permitted element is tiny or large, even if the neglect of SOC is an appropriate approximation. Such an element could turn out to be zero under the regime of spin group, providing rational guiding principles for experiments.

Last but not least, since the symmetries of spin and space degree of freedom are considered separately, spin group could provide a unified group theory for describing materials in all the four quadrants of Fig. 1. Recall that the diagnosis of degeneracy and topological phases with and without SOC has been very different because of the applications of single-valued and double-valued representations for the same symmetry operations, leading to distinct commutation relations and eigenvalues in different contexts. In the regime of spin group, the little co-group representations in the momentum space of different quadrants are naturally connected with each other by decomposing the subduced projective representation of the anti-unitary parent group, as shown in the hierarchy relationship of Fig. 2(i). To conclude, spin group serves as a bridge to connect the seemingly independent descriptions based on nonmagnetic and magnetic groups and paves a new avenue for understanding the emergent properties of magnetic quantum materials.

*Note added:* Recently, we are aware of some other works with the subject of spin group symmetry [97-100]. In [97], the magnon band structures with nodal point/line/volume protected by spin group symmetry are illustrated via Kitaev-Heisenberg models. Ref. [98] predicts a new type of collinear antiferromagnet with spin splitting via spin group symmetry analysis. Ref. [99] derives symmetry invariants of all nontrivial SPGs, indicating new excitations in the framework of spin group. Ref. [100] predicts a kind of eightfold degenerate fermion in two-dimensional antiferromagnet protected by spin group symmetry.

**Acknowledgments**

We thank Chen Fang, Zhida Song, and Zhi Wang for helpful discussions. This work was supported by National Key R&D Program of China under Grant No. 2020YFA0308900, the National Natural Science Foundation of China under Grant No. 11874195, 11834006 and 12188101, Guangdong Provincial Key Laboratory for Computational Science and Material Design under Grant No. 2019B030301001, the Science, Technology, and Innovation Commission of Shenzhen Municipality (No. ZDSYS20190902092905285) and Center for Computational Science and Engineering of Southern University of Science and Technology. X.W. also acknowledges the support from the Tencent Foundation through the XPLORER PRIZE.

*An Open-Source Software Package for Novel Topological Materials*, Comput. Phys. Commun. **224**, 405 (2018).

TABLE I. Spin space group (SSG) symmetries supporting quasi-Kramers degeneracy. The SSG symmetries, the momenta with protected 2-fold degeneracy, the surfaces that maintain the corresponding symmetry, and the possible surface states with various nodal structures are listed. Square of some operators: $T^2 = -1$, $U_n(\pi)^2 = -1$, $m_{[001]}^2 = 1$, ${\tau_{1/2}^z}^2 = 1\,(-1)$ for $k_z = 0\,(\pi)$.

| SSG symmetry | Momenta with protected 2-fold degeneracy | Surfaces with the symmetry | Possible surface states |
|---|---|---|---|
| $\{T\|\|E\|\tau_{z/2}\}$ | TRIM within $k_z = 0$ plane | $(xy0)$ | Dirac point at $(0,0)$ or $(\pi, 0)$ |
| $\{TU_z(\pi)\|\|m_{[001]}\|\tau_{x/2}\}$ | $(\pi, 0, k_z)$ and $(\pi, \pi, k_z)$ lines | $(010)$ | Dirac nodal line at $k_x = \pi$ |
| $\{T\|\|C_z(\pi)\|0\}$ | $k_z = 0$ and $k_z = \pi$ planes | $(001)$ | Possible double Dirac point |
| $\{T\|\|m_{[001]}\|0\}$ | $(0,0,k_z), (0,\pi,k_z)$, $(\pi,0,k_z)$ and $(\pi,\pi,k_z)$ lines | $(xy0)$ | Dirac nodal line at $k_x = 0$ or $k_x = \pi$ |
| $\{T\|\|m_{[001]}\|\tau_{x/2}\}$ | $(0,0,k_z)$ and $(0,\pi,k_z)$ lines | $(010)$ | Dirac nodal line at $k_x = 0$ |
| $\{TU_n(\pi)\|\|m_{[001]}\|\tau_{x/2}\}$ | $(\pi,0,k_z)$ and $(\pi,\pi,k_z)$ lines | $(010)$ | Dirac nodal line at $k_x = \pi$ |
| $\{TU_n(\pi)\|\|E\|\tau_{z/2}\}$ | TRIM within $k_z = \pi$ plane | $(xy0)$ | Dirac point at $(0,\pi)$ or $(\pi,\pi)$ |

To facilitate implementing spin group symmetry in tight-binding models obtained from atomic orbitals $[|\varphi_{R,i,\mu}, \alpha\rangle]$ ($R$ – coordinates of unit cells; $i$ – label of atomic sites; $\mu$ – label of orbitals at a site; $\alpha$ – label of spin), we write general matrix representations of the symmetry operators with the basis of Bloch wavefunctions Fourier-transformed from $|\varphi_{R,i,\mu}, \alpha\rangle$, defined as $|\psi_{k,i,\mu}, \alpha\rangle \equiv \sum_R e^{ik\cdot(R+r_i)}|\varphi_{R,i,\mu}, \alpha\rangle$ [67]:

$R(T) = -i\sigma_y K; R(U_n(\pi)) = e^{-i\pi\, n\cdot\frac{\sigma}{2}}; R_k(\tau_{n/2}) = e^{-i\frac{n\cdot k}{2}}\rho_x;$

$R(C_n(\pi)) = \oplus_w[\,M^w(C_n(\pi))\otimes[\oplus_{N=1,2,\ldots}[\oplus_{l=0}^N[e^{-i\pi\, n\cdot L^l}]]]];$

$R(m_n) = \oplus_w[\,M^w(m_n)\otimes[\oplus_{N=1,2,\ldots}[\oplus_{l=0}^N[e^{-i\pi\, n\cdot L^l}R_r^l(I)]]]];$

$w$ – index of Wyckoff positions in a crystal; $N$ – principal quantum number; $l$ – azimuthal quantum number; $\rho_x$ – the first Pauli matrix that switches two sublattices connected by $\tau_{n/2}$. $M^w(C_n(\pi))$ and $M^w(m_n)$ are representation matrices of $C_n(\pi)$ and $m_n$ with the basis of a set of atoms labeled by a Wyckoff position. $L^l$ and $R_r^l(I)$ in $R(C_n(\pi))/R(m_n)$ separately stands for the representation matrix of angular momentum and inversion operator with the basis of a set of Wannier functions of azimuthal quantum number $l$ (possibly from different atomic sites), connected by $C_n(\pi)/m_n$ symmetry. Note: since a symmetry operation could transform $|\psi_{\boldsymbol{k},i,\mu},\alpha\rangle$ into $|\psi_{\boldsymbol{k}+\boldsymbol{K},j,\nu},\beta\rangle$ where $\boldsymbol{K}$ is a reciprocal lattice vector, there might be a phase factor in $M^w$ if we consider $|\psi_{\boldsymbol{k},j,\nu},\beta\rangle$ rather than $|\psi_{\boldsymbol{k}+\boldsymbol{K},j,\nu},\beta\rangle$ as the basis vector.

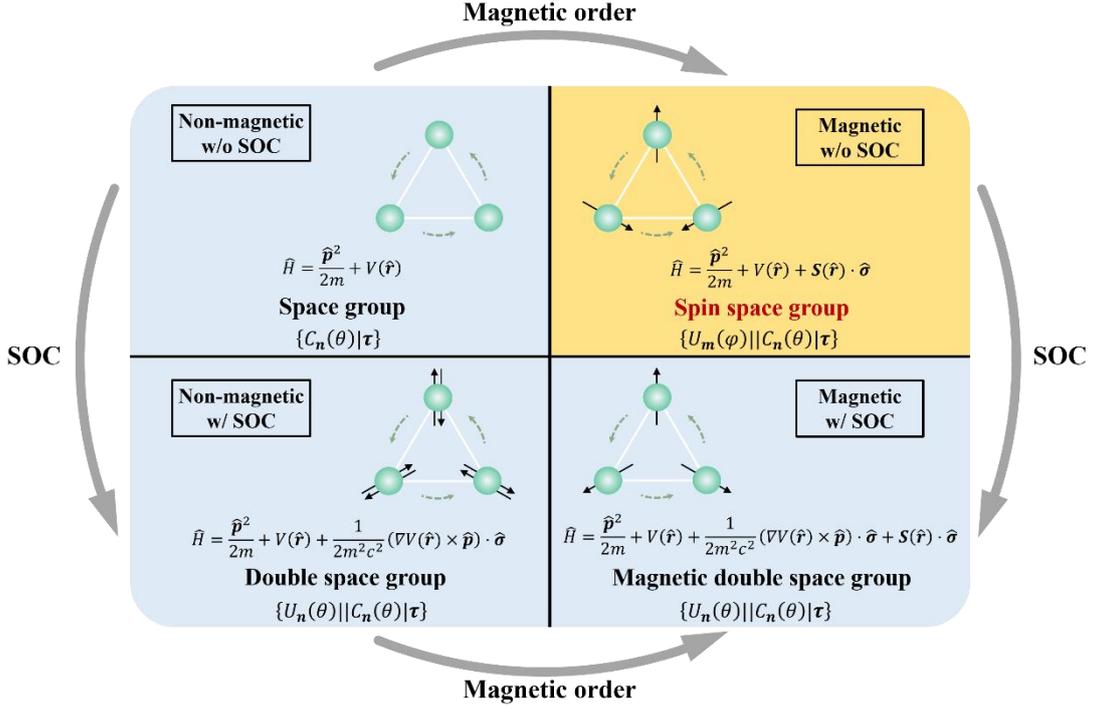

FIG. 1. Four-quadrant diagram describing the symmetry of solids with/without magnetic order and/or SOC. The general steady-state Hamiltonians, space groups, and their representative group elements are shown for each quadrant. Compared with the conventional crystallographic groups, the key characteristic of spin group is the partial decoupling between spatial rotation $C_n(\theta)$ and spin rotation $U_m(\varphi)$, where $\boldsymbol{m}$ and $\boldsymbol{n}$ denote the rotation axes, and the real scalars $\varphi$ and $\theta$ are the rotation angles. For the materials with SOC, i.e., the quadrant III and IV, spatial and spin rotations are completely locked. For example, a spatial rotation by $2\pi/3$ requires a simultaneous spin rotation by $2\pi/3$ along the same axis. For the materials without SOC, the spin and spatial rotations are completely or partially decoupled, which implies that one symmetry operation could be composed of a spin and a spatial rotation with different rotation axes and angles. For the nonmagnetic case (quadrant II), we could either consider spatial rotation only or add a totally unconstrained spin rotation, which constitutes a SO(3) group for spin. For magnetic cases (quadrant I), spin rotation is constrained by the magnetic orders of the system, which allows more operations that are disallowed by SOC but less than the full SO(3) group. The schematic plot in quadrant I shows that, for a specific magnetic order, we can have a symmetry operation that is composed of a spatial rotation of $2\pi/3$ and a spin rotation of $4\pi/3$ along the same axis. Such operations are written as $\{U_m(\varphi)||C_n(\theta)|\tau\}$ where the spatial and spin rotation axes and angles could be different.

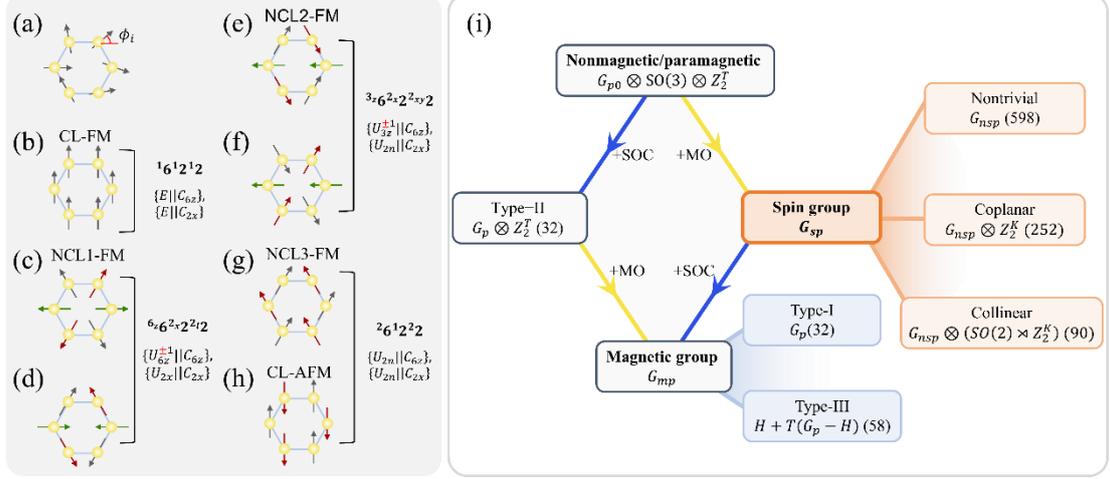

FIG. 2. Symmetries of various spin configurations within the regime of spin point group. (a) a spinful hexagonal molecule with $D_6$ spatial symmetry. (b-h) seven inequivalent spin configurations of the hexagonal molecule containing $C_{6z}$ and $C_{2x}$ spatial rotations. The corresponding nontrivial SPG symbols and generators are shown. We follow the notations of SPG in Ref. [38], with the detailed explanation provided in Appendix C. (i) symmetry hierarchy from SPG to MPG for various spin arrangements. SOC: spin-orbit coupling; MO: magnetic order. $G_{p0}$: PG for spinless system (containing spatial operations only); $G_p$: PG with complete locking between spin and spatial degrees of freedom; $G_{mp}$: MPG; $G_{sp}$: full SPG; $G_{nsp}$: nontrivial SPG.

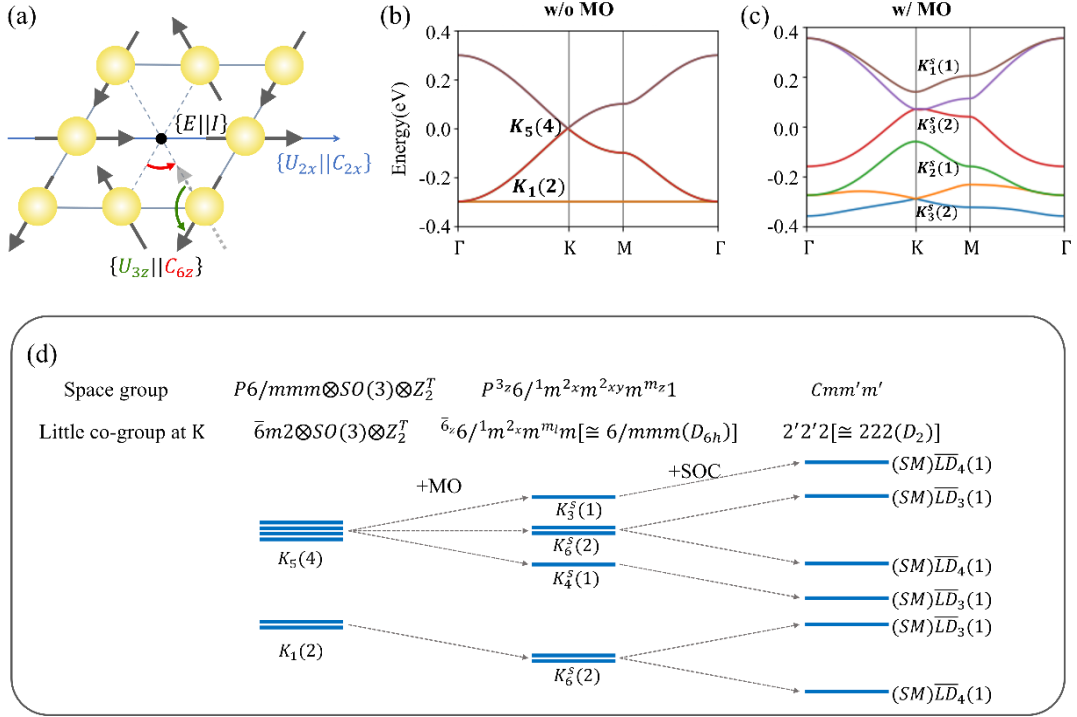

FIG. 3. (a) a spinful kagome lattice with noncolinear AFM spin configuration shown in Fig. 2 (e). (b-c) SOC-free band structure of the kagome lattice without (b) and with (c) magnetic order. The band eigenvalues at K include a 4D irreducible representation $K_4$ and a 2D irreducible representation $K_1$ if considering spin. The numbers in the parathesis represent the dimension of representations. The band eigenvalues of (c) at K include two two-fold degenerate points belonging to 2D irreducible co-representation $K_3^s$ and two nondegenerate points belonging to $K_1^s$ and $K_2^s$, respectively. (d) level splitting at K with consecutive addition of magnetic order and spin-orbit coupling. MO and SOC denote magnetic order and spin-orbit coupling, respectively. The irreducible (co-)representations of bands are shown in the diagram, with the numbers in the parathesis denoting the dimensionality (considering spin). Note: There are two K points with different little co-groups in the framework of magnetic group, and the one with higher symmetry is chosen for demonstration.

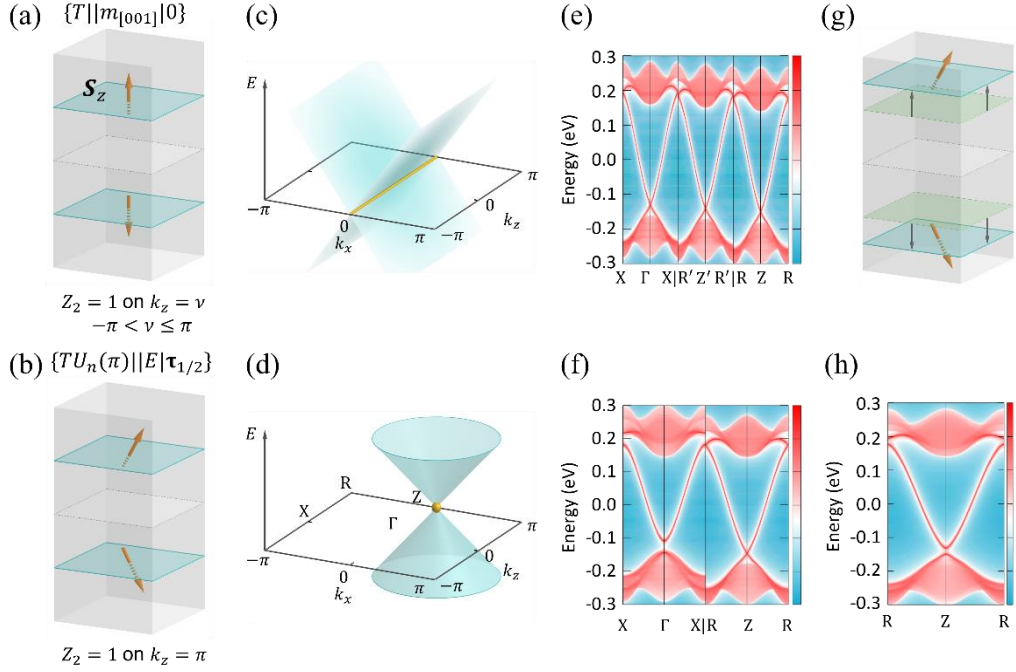

FIG. 4. $Z_2$ magnetic topological insulators protected by SSG symmetries. (a-b) a magnetic system with (a) A-type AFM structure invariant under $\{T||m_{[001]}|0\}$ and $\{TU_n(\pi)||E|\tau^z_{1/2}\}$ SSG symmetry, (b) $\{TU_n(\pi)||E|\tau^z_{1/2}\}$ SSG symmetry. (c-d) the corresponding [100] surface nodal structures in the Brillouin zone, including (c) surface nodal line and (d) surface Dirac cone. (e-f) the corresponding surface band dispersion. (g) the configuration with broken $\{TU_n(\pi)||E|\tau^z_{1/2}\}$ by the dimerization of the two layers. (h) the corresponding (100) surface band dispersion with a gapped Dirac cone.

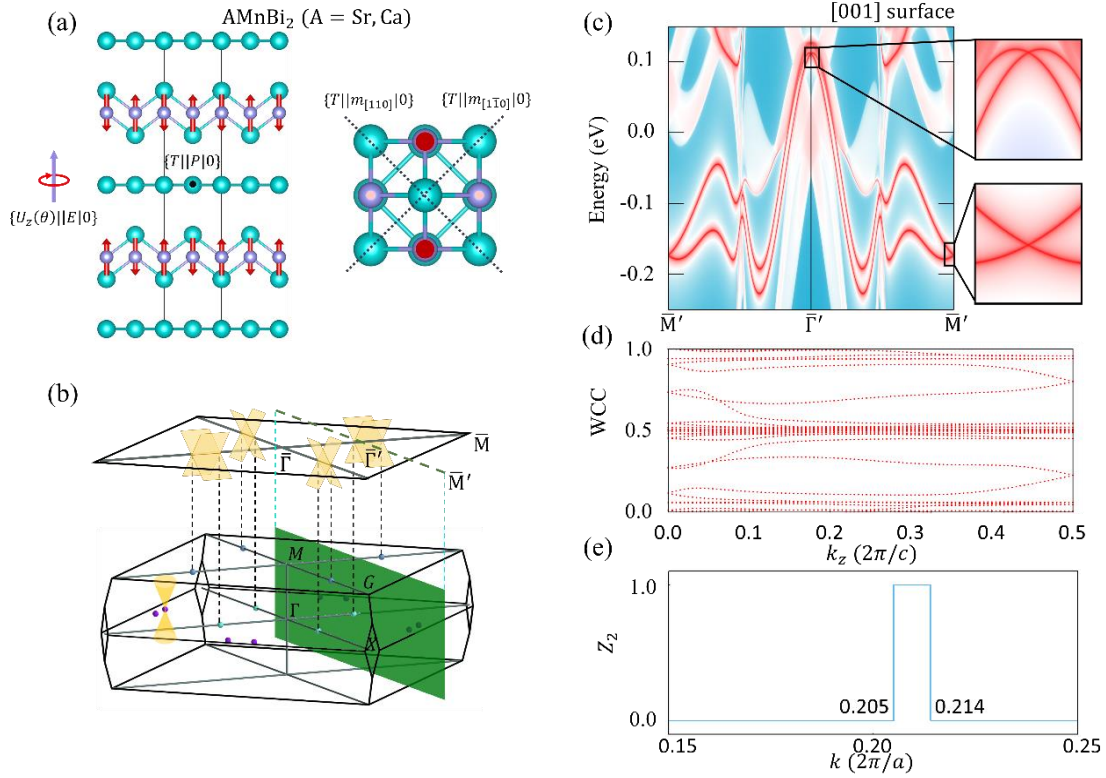

FIG. 5. $Z_2$ topological phase protected by $\{T||m_{[110]}|0\}$ symmetry in square-net materials. (a) crystal structure of AMnBi$_2$ (A = Sr, Ca). The A atoms are not plotted for clarification. (b) Locations of 16 Dirac points and the corresponding surface nodal lines of pressured SrMnBi$_2$ (with lattice parameter $c$ reduced by 10%). The equivalent Dirac points that are connected by symmetry are denoted by the same color. There are 3 types of nonequivalent Dirac points, with 8 Dirac points located at generic momenta (denoted by purple), represented by $(0.690, 0.066, 0)(\text{Å}^{-1})$; 4 Dirac points located along $\Gamma - X$ line (denoted by cyan), represented by $(0.280, 0.280, 0)$; and 4 Dirac points located along $M - G$ line (denoted by blue), represented by $(0.293, 0.293, 0.272)$. The four surface nodal lines terminate at the surface projections of the Dirac points (denoted by the dashed lines). (c) Topological nodal-line surface states of pressured SrMnBi$_2$. The $k$-path is from $\bar{\Gamma}'(0.287, 0.287)$ to $\bar{M}'(0.973, -0.399)$. (d) Wilson loop calculation in on a plane (colored by green in panel b) parallel to $m_{[110]}$. The integration along $\bar{\Gamma}' - \bar{M}'$ is calculated along $k_z$ from $k_z = 0$ to the Brillouin zone boundary $k_z = \pi/c$. (e) The transition of 2D $Z_2$ value defined at the vertical planes in the Brillouin zone parallel to $m_{[110]}$, as a function of the momentum along the [110] direction.

# APPENDIX

## A. Definition of spin group and its classifications

32 crystallographic **point groups** (**PG**s) and 230 **space groups** (**SG**s) are groups that aim to describe three-dimensional (3D) nonmagnetic crystals. For magnetic crystals, the ordered spin arrangements in periodic lattices are generally described by 122 **magnetic point groups** (**MPG**s) and 1651 **magnetic space groups** (**MSG**s). Such magnetic groups introduce the antisymmetric time-reversal operation $T$ that flips double-valued properties like spin, thus enlarging the number of crystallographic group types.

However, spin, as a vector in 3D Euclidian vector space, could have more than two values in realistic spin arrangements, the symmetry operations of which include spin rotations and spin inversion forming an orthogonal group $O(3)$ that keeps dot product of any two vectors in 3D vector space invariant. The inversion of spin is realized through $T$, for which we write the orthogonal group acting on the spin space as

$$O^s(3) = SO(3) \times Z_2^T, \tag{A1}$$

$$Z_2^T = \{E, T\}, \tag{A2}$$

$$SO(3) = \{U_n(\omega) | n = sin(\theta)cos(\varphi)\hat{x} + sin(\theta)sin(\varphi)\hat{y} + cos(\theta)\hat{z}, \theta \in (0, \pi], \varphi \in (0, 2\pi]; \omega \in (0, 2\pi]\}, \tag{A3}$$

where $U_n(\omega)$ stands for spin rotation with rotation axis $n$ and rotation angle $\omega$. Such group $O^s(3)$ is termed as the **orthogonal group of spin symmetries (OS).**

When considering a spin arrangement in real space, we must include spin operations and spatial operations at the same time. One spin arrangement could be represented by a three-component vector-valued function:

$$S(r) = (S_x(r), S_y(r), S_z(r))^T. \tag{A4}$$

Then, **spin group** $G_s$ is defined as any subgroup of the external direct product of a group with elements exerting on three-dimensional spatial coordinates (either PG or SG), denoted as $G_0$, and the orthogonal groups of spin symmetries, i.e., $O^s(3)$:

$$G_0 \otimes O^s(3) = G_0 \otimes (SO(3) \times Z_2^T). \tag{A5}$$

For $G_0$ being PG $G_{p0}$, which is written as

$$G_{p0} = \{C_n(\omega)I^m | \mathbf{n} = sin(\theta)cos(\varphi)\hat{x} + sin(\theta)sin(\varphi)\hat{y} + cos(\theta)\hat{z}, \theta \in (0,\pi],$$
$$\varphi \in (0,2\pi]; \omega \in (0,2\pi], m = 0,1\}, \tag{A6}$$

where $C_n(\omega)$ stands for a spatial rotation operation with rotation axis $\mathbf{n}$ and rotation-angle $\omega$, while $I$ stands for spatial inversion symmetry operation, every subgroup of $G_{p0} \otimes O^s(3)$ is called **spin point group** (**SPG**) $G_{sp}$, operations of which are denoted as

$$\{g_s||g\}, \tag{A7}$$

with $g_s \in O^s(3)$ and $g \in G_{p0}$. And

$$\{g_s||g\}\{g_s'||g'\} = \{g_s g_s'||gg'\}, \tag{A8}$$

$$\{g_s||g\}^{-1} = \{g_s^{-1}||g^{-1}\}. \tag{A9}$$

We define $g_s$ and $g$ acting on the spin space and coordinate space, respectively:

$$g_s: \mathbf{S} \to R(g_s)\mathbf{S}, \tag{A10}$$

$$g: \mathbf{r} \to R(g)\mathbf{r}, \tag{A11}$$

where $R(g_s)$ and $R(g)$ is the representation matrix of $g_s$ and $g$ in 3D Euclidian space, respectively. Then, for the $\mathbf{r}$-dependent spin arrangement $\mathbf{S}(\mathbf{r})$, we have

$$g_s: \mathbf{S}(\mathbf{r}) \to R(g_s)\mathbf{S}(\mathbf{r}), \tag{A12}$$

$$g: \mathbf{S}(\mathbf{r}) \to \mathbf{S}(R(g^{-1})\mathbf{r}). \tag{A13}$$

Thus, it is natural to define the action of SPG operations on the spin arrangement as

$$\{g_s||g\}: \mathbf{S}(\mathbf{r}) \to R(g_s)\mathbf{S}(R(g^{-1})\mathbf{r}). \tag{A14}$$

If we are interested in the symmetry groups of realistic molecule or crystal, we must consider another scalar-valued function $f(\mathbf{r})$ which stands for the electric potential function contributed by atomic nuclei. Since spin rotation has no influence on electric potential, the action of spin point group is defined as:

$$\{g_s||g\}: f(\mathbf{r}) \to f(R(g^{-1})\mathbf{r}). \tag{A15}$$

Note: For $g_s = U_n(\omega)/g = C_n(\omega)$, we denoted $R(U_n(\omega)) = R_n(\omega) / R(C_n(\omega)) = R_n(\omega)$, with $R_n(\omega)$ being the rotation matrix in 3D Euclidian space with rotation axis and rotation direction represented by $\mathbf{n}$ and rotation angle represented by

$\omega$. For $g_s = T/g = I$, we denote $R(T) = -\mathbb{I}/R(I) = -\mathbb{I}$, with $\mathbb{I}$ being the identity matrix.

For $G$ being space group $G_{s0}$, then every subgroup of $G_{s0} \otimes O^s(3)$ is called a **spin space group** (**SSG**) $G_{ss}$. We can further write SSG operations as

$$\{g_s||g|t\}, \tag{A16}$$

where $\{g|t\} \in G$ with $g$ denoting PG operation and $t$ is a three-component real vector denoting translation operation. We have

$$\{g_s||g|t\}\{g_s'||g'|t'\} = \{g_s g_s'||gg'|gt' + t\}, \tag{A17}$$

$$\{g_s||g|t\}^{-1} = \{g_s^{-1}||g^{-1}| - R(g^{-1})t\}. \tag{A18}$$

The actions of $\{g_s||g|t\}$ on spin arrangement and scalar potential satisfy

$$\{g_s||g|t\}: S(r) \to R(g_s)S(R(g^{-1})(r - t)), \tag{A19}$$

$$\{g_s||g|t\}: f(r) \to f(R(g^{-1})(r - t)). \tag{A20}$$

MSGs and MPGs include antisymmetry (or time-reversal) operation in addition to spatial symmetry operations while neglecting spin rotation operations. Thus, they are incomplete in a sense for describing the full symmetry of a general spin arrangement. When spin-orbit coupling is included, MSGs and MPGs are accurate for describing symmetry of Hamiltonian in physics because spin and lattice degrees of freedom must rotate synchronously, which binds the spin rotations in spatial rotations. Hence, including only spatial rotation is enough for describing full symmetry. However, when relativistic spin-orbit coupling is negligible, spin rotations and spatial rotations have to be considered separately, for which spin groups should be applied.

Ref. [101] shows that every spin group $G_s$ (either SPG $G_{sp}$ or SSG $G_{ss}$) can be written as a direct product of a so-called **spin-only group** $G_{so}$ and a **nontrivial spin group** $G_{ns}$. Spin-only group stands for the group formed by pure spin operations $\{g_s||E|0\}$ (or $\{g_s||E\}$), while the nontrivial spin group stands for the group that contain no pure spin operations, i.e., all of the group elements contain spatial operations except the identity.

In Appendix B, we analyze all possible Spin-only groups for different types of spin arrangements. In Appendix C, we comprehensively develop all the possible combinations between nontrivial SPGs and spin-only PGs to provide a full description of the possible SPGs.

## B. Groups consisting of pure spin operations

A spin-only group consists of elements of the form $\{g_s||E|0\}$ (or $\{g_s||E\}$), which act on the spin configuration $S(r)$ as

$$\{g_s||E|0\}: S(r) \to R(g_s)S(r). \tag{B1}$$

To analyze the spin-only groups for different spin arrangements, we divide all of the pure spin operations into 4 types, i.e., $g_s = U_n(\omega)$ ($\omega \neq 0$), $g_s = T$, $g_s = U_n(\pi)T$ and $g_s = U_n(\omega)T$ ($\omega \neq 0 \text{ and } \omega \neq \pi$), and analyze the conditions for $S(r)$ to be invariants under them separately. (We choose specific axes of spin rotations to simplify our analysis)

**Type-1:** $g_s = U_n(\omega)$ for $\omega \neq 0$

In this case, we have $S(r) = R(U_n(\omega))S(r)$. Then, there are two types of spin arrangements that have $U_n(\omega)$ symmetry: 1) $S(r)$ is zero for any $r$, corresponding to **nonmagnetic spin arrangements**. 2) If $S(r)$ does not belong to nonmagnetic spin arrangement, $S(r)$ should be parallel to $n$, corresponding to **collinear spin arrangements**.

**Type-2:** $g_s = T$

If $g_s = T$, then the spin arrangements invariant under $\{T||E|0\}$ should satisfy $S(r) = -S(r)$, indicating nonmagnetic spin arrangements.

**Type-3:** $g_s = U_n(\pi)T$

If $n = z$, then the spin arrangements invariant under $\{U_z(\pi)T||E|0\}$ should satisfy $S(r) = \begin{pmatrix} 1 & 0 & 0 \\ 0 & 1 & 0 \\ 0 & 0 & -1 \end{pmatrix} S(r)$. Then, the spin arrangements $S(r)$ that have no $z$ component have such symmetry. We define all spin arrangements satisfying this condition that are not nonmagnetic or collinear spin arrangements as **coplanar spin arrangements**. Apparently, each spin arrangement that is nonmagnetic, collinear, or coplanar is invariant under $\{U_n(\pi)T||E|0\}$ for certain $n$. Consequently, the spin arrangements that are not coplanar, collinear or nonmagnetic are known as **noncoplanar spin arrangements**.

**Type-4:** $g_s = U_n(\omega)T$ ($\omega \neq 0$ and $\omega \neq \pi$)

If $n = z$, then the spin arrangement that is invariant under $\{U_z(\omega)T||E|0\}$ satisfy

$$S(r) = \begin{bmatrix} -Cos(\omega) & Sin(\omega) & 0 \\ -Sin(\omega) & -Cos(\omega) & 0 \\ 0 & 0 & -1 \end{bmatrix} S(r).$$ It is easy to check that such equation has no solution for $S(r)$ unless $S(r) = 0$.

In conclusion, there are 4 types of spin arrangements that have different spin-only groups. Since the spatial part is always identity in spin-only groups, we only write down the spin part of $G_{so}$ for simplicity in the following.

1. **Nonmagnetic spin arrangements**

    The spin-only group is invariant under all the pure spin operations:
    $$G_{so} = O^s(3). \tag{B2}$$

2. **Collinear spin arrangements**

    Any spin rotation along the direction of spin arrangement (e.g., $z$ direction) could leave this type of spin arrangement invariant, indicating $SO(2) \equiv \{U_z(\omega)|\omega \in (0,2\pi]\}$ spin rotation group. Furthermore, such spin arrangement is also invariant under $U_n(\pi)T$, where $n$ could be any direction perpendicular to $z$. We then have a binary spin-only group defined as $Z_2^K \equiv \{E, U_n(\pi)T\}$. Then the full spin-only group is the internal semidirect product of $SO(2)$ and $Z_2^K$, i.e.,
    $$G_{so} = SO(2) \rtimes Z_2^K. \tag{B3}$$

    Note that the internal semidirect product is because $SO(2)$ is a normal subgroup of $G_{so}$ while $Z_2^K$ is not.

3. **Coplanar spin arrangements**

    From the discussion above, the spin-only group of such spin arrangement is
    $$G_{so} = Z_2^K \equiv \{E, U_n(\pi)T\}, \tag{B4}$$
    where $n$ denotes the direction perpendicular to the plane of the spin arrangements.

4. **Noncoplanar spin arrangements**

    The spin-only group of this type has only the identity element
    $$G_{so} = \{E\}. \tag{B5}$$

## C. Full crystallographic spin point groups for collinear and coplanar spin arrangements

Crystallographic SPGs are SPGs being subgroups of $G_{p0} \otimes O^s(3)$ where $G_{p0}$ is one of the 32 crystallographic PGs. Construction of nontrivial SPGs from 32 PGs is similar to obtaining the MPGs. However, the orthogonal group of spin symmetries, $SO(3) \times Z_2^T$, has an infinite number of operations, the introduction of which into crystallographic PGs requires us to find all of the normal subgroups of the 32 PGs and find all of the groups that are subgroups of $SO(3) \times Z_2^T$ and isomorphic to the corresponding quotient groups. Ref. [37] shows that all the nontrivial SPGs can be obtained in the above approach if we find all normal subgroups of the 32 PGs, construct all quotient groups from these normal subgroups, and find all subgroups of $SO(3) \times Z_2^T$ that are isomorphic to the quotient groups through all possible isomorphic relations. By applying such procedure, Ref. [38] obtains 598 types of nontrivial SPGs, with two groups defined to belong to the same type if they are conjugate subgroups of the direct product of general linear group in spin space and affine group in physical space, i.e., $GL(3) \otimes GIL(3)$, which are sufficient to describe noncoplanar spin arrangements according to the discussion in Appendix B.

However, for collinear and coplanar spin arrangements, there are pure spin operations, i.e., we have to consider SPGs that are products of nontrivial SPGs and spin-only PGs, i.e., $G_{sp} = G_{nsp}G_{sop}$. Such kind of products could always be written as semidirect product $G_{nsp} \ltimes G_{sop}$ because $G_{sop}$ is always a normal subgroup of $G_{nsp}G_{sop}$. This could also be written as a direct product $G_{nsp} \times G_{sop}$ by proper selection of $G_{nsp}$ such that $G_{nsp}$ is also a normal subgroup of $G_{nsp}G_{sop}$. Thus, to classify the point groups of the form $G_{nsp} \times G_{sop}$, we have to find all the types of nontrivial SPG ($G_{nsp}$) that could perform internal direct product with spin-only groups $G_{sop}$. Since 598 types of nontrivial SPG are complete, and any SPGs can be written as the direct product of nontrivial SPGs and spin-only PGs, such classification should lead to a complete set of SPGs for describing coplanar and collinear spin arrangements. (We neglect nonmagnetic spin arrangements because they obviously have symmetry groups of the form $G_0 \otimes O^s(3)$ with $G_0$ being one of the PGs or SGs). We conduct such classification in the following 2-step procedure:

**Step 1:** We find all of the types of nontrivial SPG that allow both $G_{nsp}$ and $G_{sop}$ are invariant under each other, i.e., satisfy $g^{-1}G_{nsp}g = G_{nsp}$ for all $g \in G_{sop}$ and $h^{-1}G_{sop}h = G_{sop}$ for all $h \in G_{nsp}$, and then determine the corresponding $G_{nsp}G_{sop}$s (one nontrivial SPG might correspond to several full SPGs). Since both $G_{nsp}$ and $G_{tsp}$ are subgroups of $G_{p0} \otimes O^s(3)$, and $G_{nsp} \cap G_{sop} = \{\{E||E\}\}$, The condition, $g^{-1}G_{nsp}g = G_{nsp}$ for all $g \in G_{sop}$ and $h^{-1}G_{sop}h = G_{sop}$ for all $h \in G_{nsp}$, implies that $G_{nsp}G_{sop}$ is a group and $G_{nsp}G_{sop} = G_{nsp} \times G_{sop}$. Then, we get all types of full SPGs represented by $G_{nsp} \times G_{sop}$.

**Step 2:** We consider full SPGs obtained from Step 1, represented as groups $G_{nsp}^1 \times G_{sop}, G_{nsp}^2 \times G_{sop}, \ldots, G_{nsp}^n \times G_{sop}$, with $G_{nsp}^1, G_{nsp}^2, \ldots$ and $G_{nsp}^n$ belonging to different types of nontrivial SPGs, but with $G_{nsp}^1 \times G_{sop}, G_{nsp}^2 \times G_{sop}, \ldots$ and $G_{nsp}^n \times G_{sop}$ actually belonging to the same type of full SPGs. Then, we choose one of the $G_{nsp}^i \times G_{sop}$ to represent this full SPG. Or in other words, eliminate multiple counting of the equivalent types of full SPGs.

As discussed in Appendix B, the full SPGs for coplanar spin arrangements can be written as $G_{nsp} \times Z_2^K$, while the full SPGs for collinear spin arrangements could be described by $G_{nsp} \times (Z_2^K \ltimes SO(2))$. Next, we separately classify full SPGs for coplanar spin arrangements and for collinear spin arrangements.

**C1. Classification of full SPGs for coplanar spin arrangements**

For coplanar spin arrangements, we have $G_{sop} = Z_2^K$. Note that we do not consider the relative directions of spin rotation axis and space rotation axis in the following discussion because the variation of relative direction of spin rotation axis and space rotation axis will not give rise to the different types of spin group.

**Step 1**

The step 1 outlined above implies that if we write $Z_2^K \equiv \{\{E||E\}, \{TU_n(\pi)||E\}\}$,

the rotation axis $\boldsymbol{n}$ should be either parallel or perpendicular to every spin rotation axis of the spin rotation part of $G_{nsp}$, i.e., $G_{nsp}^S$. Otherwise, the condition that $G_{sop} = Z_2^K$ should be invariant under $G_{nsp}$ will not be satisfied. Thus, this step excludes the nontrivial SPGs whose spin part $G_p^S$ are polyhedral group, including $T, T_h, T_d, O$ and $O_h$, because we cannot find a direction that is either parallel or perpendicular to all the rotation axis. Thus, the options left for the spin part $G_{nsp}^S$ are 27 axial groups: $C_n(n = 1,2,3,4,6)$, $D_n(n = 2,3,4,6)$, $S_n(n = 2,4,6))$ ($S_2 = C_i, S_6 = C_{3i}$), $C_{nh}(n = 1,2,3,4,6)$ ($C_{1h} = C_s = C_{1v}$), $D_{nh}(n = 2,3,4,6)$, $C_{nv}(n = 2,3,4,6)$, and $D_{nd}(n = 2,3)$. For $G_{nsp}^S$ being different groups, the direction $\boldsymbol{n}$ is constrained differently in order that the condition that $g^{-1} Z_2^K g = Z_2^K$ for all $g \in G_{nsp}$ is satisfied. We classify the ways $\boldsymbol{n}$ is constrained into the following 5 cases.

**Case 1:** For $G_{nsp}^S$ being $C_1$ or $S_2$ ($C_i$), $\boldsymbol{n}$ is not constrained.

**Case 2:** For $G_{nsp}^S$ being $C_{1h} = \{\{E||E\}, \{TU_{\hat{z}}(\pi)|E\}\}$, $C_2 = \{\{E||E\}, \{U_z(\pi)|E\}\}$ or $C_{2h} = \{\{E||E\}, \{U_z(\pi)|E\}, \{TU_z(\pi)|E\}, \{T|E\}\}$, there is only one two-fold spin rotation $U_z(\pi)$ in $G_{nsp}^S$. Thus $\boldsymbol{n}$ which could be either parallel or perpendicular to $\boldsymbol{z}$. (When $\boldsymbol{n}$ is perpendicular to $\boldsymbol{z}$, then groups corresponding to different $\boldsymbol{n}$ belong to the same types up to conjugate transformations.)

**Case 3:** For $G_{nsp}^S$ being the groups that have rotations of order larger than 2, $\boldsymbol{n}$ should be parallel to the principal axis of $G_{nsp}^S$.

**Case 4:** For $G_{nsp}^S$ being $D_2$ or $D_{2h}$, $\boldsymbol{n}$ should be parallel to one of the 2-fold spin rotation axes.

**Case 5:** For $G_{nsp}^S$ being $C_{2v}$, $\boldsymbol{n}$ should either be perpendicular to one mirror or parallel to the two-fold rotation axis.

It is easy to see that for the 5 cases, the condition that $g^{-1} G_{nsp} g = G_{nsp}$ for all $g \in Z_2^K$ is also satisfied. Then, we get all of the types of SPGs that can be written as $G_{nsp} \times Z_2^K$ with some types possibly being identical.

**Step 2**

Some types obtained from step 1 are identical because the operations that implicitly contain $T$ in $G_{nsp}$ could always be changed to the product of $\{TU_n(\pi)||E\}$ with those operations, for the full group being $G_{nsp} \times Z_2^K$. That is to say if we have a spin point group $G_{nsp}^1 \times Z_2^K$ with $G_{nsp}^1$ been a nontrivial SPG that could be written as the form $G_{nsp}^1 = H + \{T||E\}(G_{nsp}^1 - H)$ with $H$ being the subgroup of $G_{nsp}^1$ of order 2 that does not contain $T$, then, we have:

$$G_{nsp}^1 \times Z_2^K$$
$$= (H + \{T||E\}(G_{nsp}^1 - H)) \times \{\{E||E\}, \{TU_n(\pi)||E\}\}$$
$$= (H + \{U_n(\pi)||E\}(G_{nsp}^1 - H)) \times \{\{E||E\}, \{TU_n(\pi)||E\}\}$$
$$= G_{nsp}^2 \times Z_2^K, \tag{C1}$$

with $G_{nsp}^2$ containing no $T$. This implies that $G_{nsp}$ in all of the SPGs of the form $G_{nsp} \times Z_2^K$ can be chosen such that the $G_{nsp}^S$ corresponding to $G_{nsp}$ is formed by pure spin rotations.

Furthermore, it is obvious that two SPGs, $G_{nsp}^A \times Z_2^K$ and $G_{nsp}^B \times Z_2^K$, with $G_{nsp}^A$ and $G_{nsp}^B$ being different types and containing no $T$, should be different types of SPG. Thus, we can use all of the nontrivial SPGs $G_{nsp}$ that have spin part $G_{nsp}^S$ being 9 axial PGs of pure spin rotation $C_n (n = 1, 2, 3, 4, 6)$ and $D_n (n = 2, 3, 4, 6)$ to construct all types of SPGs of the form $G_{nsp} \times Z_2^K$ to avoid multiple counting of the same type of group. Then there are 4 cases left:

**Case 1:** For $G_{nsp}^S = C_1$, $\boldsymbol{n}$ could be random directions which result in the same types of SPG.

**Case 2:** For $G_{nsp}^S = C_2 = \{\{E||E\}, \{U_z(\pi)|E\}\}$, $\boldsymbol{n}$ could be either parallel or perpendicular to the $z$-axis.

**Case 3:** For $G_{nsp}^S = D_2 = \{\{E||E\}, \{U_x(\pi)|E\}, \{U_y(\pi)|E\}, \{U_z(\pi)|E\}\}$, $\boldsymbol{n}$ could be parallel to one of the 3 two-fold rotation axes. These 3 two-fold rotation axes are equivalent for $G_{nsp}^S$. But they are not necessarily equivalent for the whole nontrivial SPG $G_{nsp}$. Thus, the groups which separately have $\boldsymbol{n}$ parallel to the $\boldsymbol{x}$, $\boldsymbol{y}$ and $\boldsymbol{z}$-axis could belong to the same types of SPG or different types of SPG.

**Case 4:** For $G_{nsp}^s$ being one of the left 6 groups $C_n(n = 3, 4, 6)$ and $D_n(n = 3, 4, 6)$, $\boldsymbol{n}$ has to be parallel to the principal axis $\boldsymbol{z}$.

Thus, we can classify the types of full SPGs of the form $G_{nsp} \times Z_2^K$ into 11 categories, shown in Table C1. Finally, we get 252 types of crystallographic SPGs of the form $G_{nsp} \times Z_2^K$. These SPGs are listed in Table S1-S11 of Supplementary Materials [64].

TABLE C1. Classifications of the SPGs of the form $G_{nsp} \times Z_2^K$, with $Z_2^K \equiv \{\{E||E\}, \{TU_{\boldsymbol{n}}(\pi)||E\}\}$ and the corresponding $G_{nsp}^s$ being one of the 9 groups $C_n(n = 1,2,3,4,6)$, $D_n(n = 2,3,4,6)$. And the principal axes of $G_{nsp}^s$, if exists. are assumed to be along the $\boldsymbol{z}$ direction.

|  | $G_{nsp}^s$ | Direction of $\boldsymbol{n}$ | Number of types |
|---|---|---|---|
| Category-I | $C_1$ | random | 32 |
| Category-II | $C_2$ | parallel to the $\boldsymbol{z}$-axis | 58 |
| Category-III | $C_2$ | perpendicular to the $\boldsymbol{x}$ −axis | 58 |
| Category-IV | $C_3$ | parallel to the $\boldsymbol{z}$ −axis | 6 |
| Category-V | $C_4$ | parallel to the $\boldsymbol{z}$ −axis | 4 |
| Category-VI | $C_6$ | parallel to the $\boldsymbol{z}$ −axis | 7 |
| Category-VII | $D_2$ | parallel to $\boldsymbol{x}, \boldsymbol{y}$ or $\boldsymbol{z}$ −axis(when the three directions lead to identical types) | 2 |
| Category-VIII | $D_2$ | parallel to the $\boldsymbol{x}, \boldsymbol{y}$ or $\boldsymbol{z}$-axis (when the three directions lead to different types) | 62 |
| Category-IX | $D_3$ | parallel to the $\boldsymbol{z}$ −axis | 10 |
| Category-X | $D_4$ | parallel to the $\boldsymbol{z}$ −axis | 5 |

| | | parallel to the | |
|---|---|---|---|
| Category-XI | $D_6$ | $z-$axis | 8 |
| Total number of types: 252 | | | |

## C2. Classification of full SPGs for collinear spin arrangements

For collinear spin arrangements, we then consider the form $G_{nsp} \times (Z_2^K \ltimes SO(2))$.

**Step 1**

We firstly write $SO(2)$ as $SO(2) = \{\{U_{\boldsymbol{m}}(\omega)T||E|0\}|\omega \in (0,2\pi]\}$. Because of the presence of $SO(2) = \{\{U_{\boldsymbol{m}}(\omega)T||E|0\}|\omega \in (0,2\pi]\}$, all of the rotations contained in $G_{nsp}^S$ should have rotation axis parallel to $\boldsymbol{m}$, so that the condition that $h^{-1}G_{nsp}h = G_{nsp}$ for all $h \in Z_2^K \ltimes SO(2)$ is satisfied. Thus, $G_{nsp}^S$ should have no more than 1 rotation axis. Then, the options left for $G_{nsp}^S$ are 13 PGs: $C_n(n = 1, 2, 3, 4, 6)$, $C_{nh}(n = 1, 2, 3, 4, 6)$, $S_n(n = 2, 4, 6)$. For $G_{nsp}^S$ being these PGs, the condition that $g^{-1}[Z_2^K \ltimes SO(2)]g = Z_2^K \ltimes SO(2)$ for all $g \in G_{nsp}$ is also satisfied.

**Step 2**

Since $\boldsymbol{m}$ is parallel to the axis of rotation of $G_{nsp}^S$, and $SO(2)$ contains spin rotations along $\boldsymbol{m}$ with arbitrary rotation angle if there is a spin group of the form $G_{nsp} \times (Z_2^K \ltimes SO(2))$, then $G_{nsp}$ could be chosen such that there is no spin rotation at all in $G_{nsp}$, similar to the chosen of $G_{nsp}$ in $G_{nsp} \times Z_2^K$ such that $G_{nsp}$ contains no $T$, described in section C1.

Therefore, the options left for $G_{nsp}^S$ are $C_1$ and $S_2(= C_i = Z_2^T)$ so that there is no multiple counting of the same types of full SPGs represented by $G_{nsp} \times (Z_2^K \ltimes SO(2))$. The remaining $G_{nsp}$'s to be considered are actually 90 types, including 32 type-I MPGs and 58 type-II MPGs. Overall, there are 90 crystallographic SPGs of the form $G_{nsp} \times (Z_2^K \ltimes SO(2))$, shown in Table S12-S13 of Supplementary Materials [64].

## D. Irreducible co-representations of spin groups

For $Mn_3Sn$ Kagome lattice, we provide the specific band co-representations of the little cogroups at different momenta (K and M). The procedure is explained in the following: 1) Find the maximal unitary subgroups (MUSGs) of particular little co-groups. 2) Get the full list of irreducible representations of MUSGs from single-valued representations of their corresponding spatial part. 3) Construct the irreducible co-representations of little co-groups from those representations obtained in 2) based on the method derived by E. Wigner [102].

Table S14 and S16 list the representation matrices of all irreducible co-representations of K and M. The correspondences among the irreducible co-representations of the spin group, the single-valued representations of the corresponding nonmagnetic group, and double-valued co-representations of the corresponding magnetic group are separately listed in Table S15 and S17. Since the character tables are only defined for unitary groups, we list the character tables of MUSGs, all symmetry operations, and conjugate classes in Table S18 and S21. Since each nontrivial spin point group is isomorphic to a nonmagnetic or magnetic point group by neglecting the spin rotations, the conjugate classes of MUSGs can be directly obtained by resorting to the conjugate classes of the corresponding nonmagnetic point group.

## E. Methods of first-principles calculations

First-principles calculations of $SrMnBi_2$ and $CaMnBi_2$ are conducted within the framework of density functional theory [103,104], using projector-augmented-wave (PAW) method [105], implemented in Vienna ab initio simulation package (VASP) [106]. Pseudopotentials under the generalized gradient approximation (GGA) with the Perdew-Burke-ErnZerhof (PBE) formalism [107,108] are used. The ground state is obtained from a self-consistent calculation with an energy cut off of 500 eV and $5\times5\times5$ ($7\times5\times5$) Monkhorst-Pack grid for the symmetrized primitive cells of $SrMnBi_2$ ($CaMnBi_2$). Due to the local magnetic moments contributed from $3d$ electrons in Mn atoms, GGA+U approach within the Dudarev scheme is applied with U = 3 eV on $3d$ orbitals of Mn. To obtain the topological properties of $SrMnBi_2$ and $CaMnBi_2$, e.g., Dirac points, surface states, Wilson loop, etc., we construct tight-binding Hamiltonians based on maximally localized Wannier functions [109,110] of Sr-$4d$, Mn-$3d$, Bi-$6p$ for $SrMnBi_2$ and Ca-$3d$, Mn-$3d$, Bi-$6p$ in $CaMnBi_2$. The iterative Green's function implemented in WannierTools package is used for surface states calculations [111].

In our calculations, the lattice parameters and atomic positions of $SrMnBi_2$ and $CaMnBi_2$ are chosen from the experimental values [71] except the $z$-direction lattice parameter. To simulate the uniaxial pressure effect, we set lattice parameter $c$ of $SrMnBi_2$ to be 21.13 Å (rather than 23.13 Å) to realize the $Z_2$ topological phase in the subspaces of the Brillouin zone.